\documentclass[aps,twocolumn,showpacs,prl,superscriptaddress]{revtex4}
\usepackage{amssymb,isolatin1,graphicx,times}
\newcommand{\cbeta}{{\cal B}}

\newcommand{\muz}{\langle{\mu}\rangle}

\begin{document}
\title{Deceptive signals of phase transitions in small magnetic clusters}
\author{Heinrich Stamerjohanns}
\author{Oliver Mülken}
\affiliation{Department of Physics, Carl von Ossietzky University
Oldenburg, D-26111 Oldenburg, Germany}
\author{Peter Borrmann}
\affiliation{IBM Unternehmensberatung, Am Sandtorkai 73, 20457 Hamburg, Germany}
\date{\today}
\begin{abstract}
We present an analysis of the thermodynamic properties of small
transition metal clusters and show how the commonly used indicators of
phase transitions like peaks in the specific heat or magnetic
susceptibility can lead to deceptive interpretations of the
underlying physics. The analysis of the distribution of zeros of the
canonical partition function in the whole complex temperature plane
reveals the nature of the transition. We show
that signals in the magnetic susceptibility at positive temperatures
have their origin at zeros lying at negative temperatures.
\end{abstract}
\maketitle

Experiments on Bose-Einstein
condensation~\cite{Anderson:1995,Bradley:1995,Davis:1995} or the
experimental determination of structural, electronic and thermal
properties of clusters~\cite{Schmidt97,Schmidt98,Schmidt2001} are
prototypes of physical investigations of transitions in small systems.
Intuitively phase transitions in such systems do exist. While the
atomic structure at low temperatures is more or less rigid, at high
temperatures the atoms move resembling a liquid drop. But the theoretical
description is complicated since the thermodynamic functions of clusters
do not show singularities at the transition point. Phase changes are seen
in blurred slopes or humps. To have the physical concept of phase
transitions and their properties of bulk material in mind and apply it to
interpret the smooth thermodynamic functions for small systems, e.g.
clusters, of a given size accordingly and assign a ``first'' or ``second''
order phase transition may be inconclusive. Despite the ambiguity in these
signals, it is still reasonable to attribute an order to phase changes in
small systems because fundamental differences between the kind of the
transitions such as the existence of metastabilities for first-order
transitions still persist. Therefore, various approaches for the
classification of phase transitions in small systems have been developed
which have to coincide in the thermodynamic limit and should be
mathematically rigourous.

Gross {\sl et al.}~have proposed a microcanonical treatment, where phase
transitions are distinguished by the curvature of the entropy
$S(E)$~\cite{Gross:1997,Dgross:2000}. If $S(E)$ has a convex intruder,
i.e.~the microcanonical caloric curve $T(E)$ shows a backbending, the
phase transition is assumed to be of first order. Franzosi {\sl et
al.}~have started by investigating the topology of the potential energy
surface and established a connection between topological changes and phase
transitions~\cite{Pettini:1999,Pettini:1999a}. However, they are not able
to determine the order of the phase transition. Recently we have proposed
a classification scheme based on the distribution of zeros of the
analytically continued canonical partition function $Z(\cbeta)$, with
$\cbeta=\beta+i\tau$ ($\beta=1/T$), in the complex temperature
plane~\cite{Borr99b,Muelken:2000b,Muelken:2000,Muelken:2001}. 

The basic principle of the description of phase transitions by the zeros
of the partition function is the {\sl product theorem of Weierstrass} and
the {\sl theorem of Mittag-Leffler} which relate integral functions to
their zeros~\cite{Titchmarsh}. Applying these theorems, the canonical
partition function can be written as
\begin{eqnarray}
Z(\beta) &=& \left(\frac{1}{2\pi\beta}\right)^{3N/2} \ \int {\rm d}^{3N}q \
\exp \left[-\beta V(q) \right] \\
&=& \left( \frac{1}{2 \pi \beta}\right)^{3N/2}
 \prod_{k=-M}^M \left( 1-\frac{\beta}{\cbeta_k} \right) \exp\left(
\frac{\beta}{\cbeta_k}\right). \label{partfunc}
\end{eqnarray}

\begin{figure}[h]
\includegraphics[clip=,width=8cm]{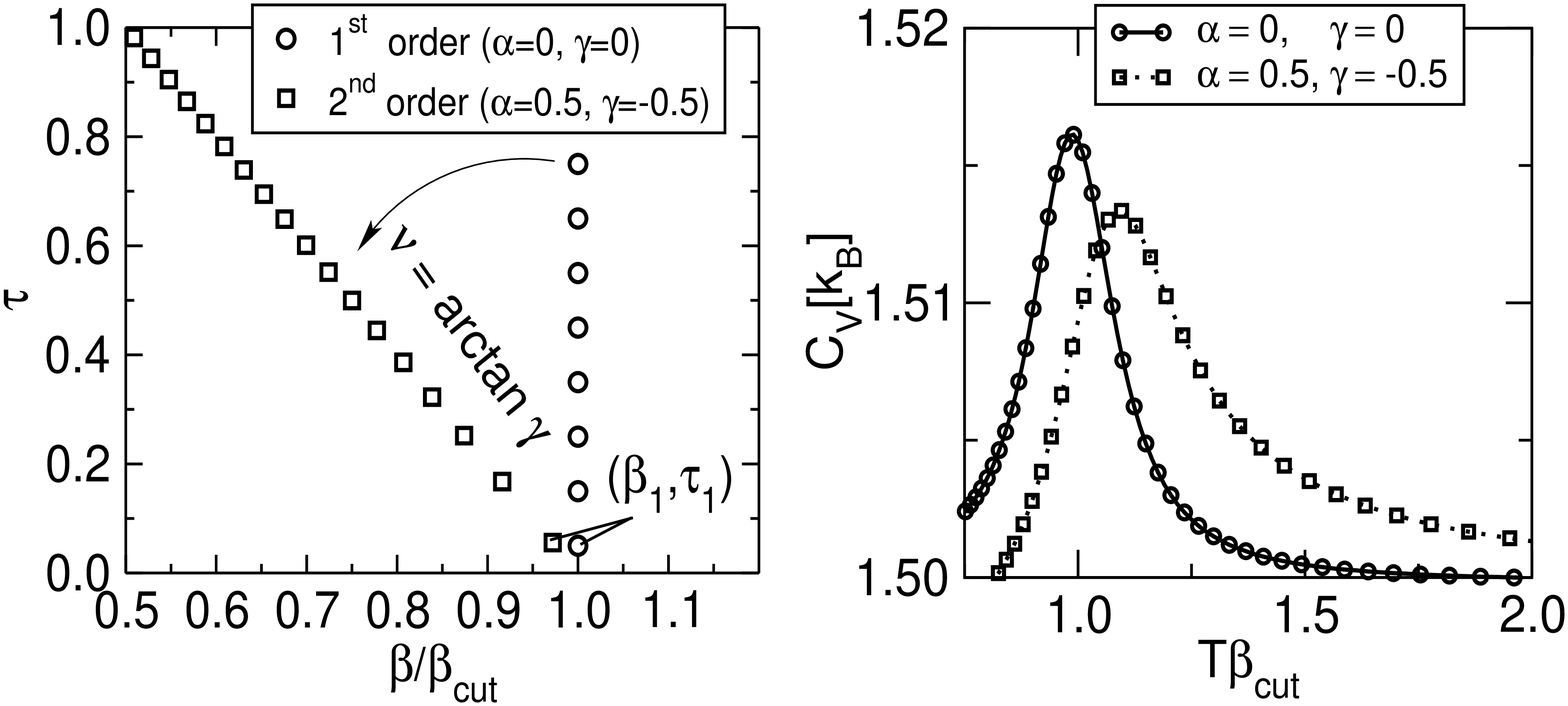}
\caption{Examples of distributions of zeros for 1$^{\rm st}$ and 2$^{\rm nd}$ order phase transitions along with the corresponding specific heat,
calculated as functions of the zeros.}
\label{fig:zeros}
\end{figure}

We assume the zeros to lie on a line with a density
$\phi(\tau)\sim\tau^\alpha$ and to have a crossing angle $\nu$ with the
plummet on the real temperature axis ($\gamma = \tan\nu$). Together with
the imaginary part $\tau_1$ of the first zero $\cbeta_1$ this leads to a
distinct classification of phase transitions in small systems. For zeros
perpendicular to the real axis with equal or increasing spacing,
i.e., $\alpha\leq0$ and
$\gamma=0$, the transition is of first order, for
$0<\alpha<1$ and arbitrary $\gamma$ as well as for $\gamma\neq0$ and
$\alpha\leq0$ of second order, see Fig.~\ref{fig:zeros}. The imaginary part
$\tau_1$ reflects the ``discreteness'' of the system. Thus, in the
thermodynamic limit we have $\tau_1 \to 0$ and our scheme coincides with
the scheme given by Grossmann and coworkers~\cite{Gross:1967}.

We utilize small magnetic clusters exposed to an external magnetic field
in order to show how the common treatment of phase transitions in small
systems like the identification by humps of response functions eventually
leads to misinterpretations of physical properties.

Metal clusters have the intriguing property that they occur as different
isomers with almost equal ground state energies but very different
magnetic moments and different geometries~\cite{bdht96,bsh+99}. For
simplicity, we consider in our model two isomers with magnetic moments
$\mu_{1}=1\mu_{\rm B}$ and $\mu_{2}=10\mu_{\rm B}$, and their ground state
energy difference $\Delta E=E_{0}(2)-E_{0}(1)=1$~meV. In the presence of an
external magnetic field $H$ pointing in $z$-direction the partition
function reads 
\begin{equation}
Z(\beta) =
\sum_{i=1}^{2} \exp[-\beta E_{0}(i)] \frac{2}{\beta \mu_{i} H} \sinh(\beta
\mu_{i} H).\label{partmag}
\end{equation}  
We have assumed equal vibrational energies.  
The two isomers can be identified by their average magnetic moment
$\muz$ which are calculated by standard differentiation of
Eq.(\ref{partmag}) with regard to the magnetic field
\begin{eqnarray}
\muz &=& \beta^{-1} \ \partial_{H}\ln Z(\beta) = \sum_i p_i \left< \mu_i
\right> \\
&=& \sum_i p_i \;[\mu_i \tanh^{-1} (\beta \mu_i H)
- 1/(\beta H)],
\end{eqnarray}
where $p_i$ is the probability of finding isomer $i$. 

\begin{figure}[h]
\includegraphics[clip=,width=8cm]{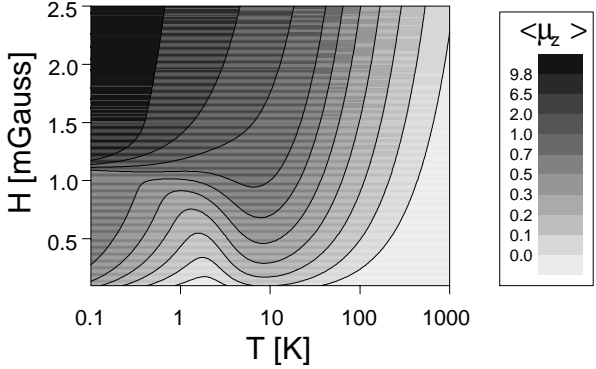}
\caption{Contour plot af the average magnetic moment $\muz$ versus
temperature $T$ and magnetic field $H$. Note the non-linear scale.}
\label{fig:mu-cont}
\end{figure}

This system is driven by two effects, the entropy increase due to thermal
excitation and the alignment of the magnetic moments along the magnetic
field. For low temperatures, there is a transition from $\muz\simeq 1$ to
$\muz\simeq 10$ at about $1.1~$mGauss, as shown in Fig.~\ref{fig:mu-cont}.
At higher temperatures the magnetic field does not align the magnetic
moments along the field resulting in a general decrease of $\muz$. 

\begin{figure}[h]
\includegraphics[clip=,width=7.5cm]{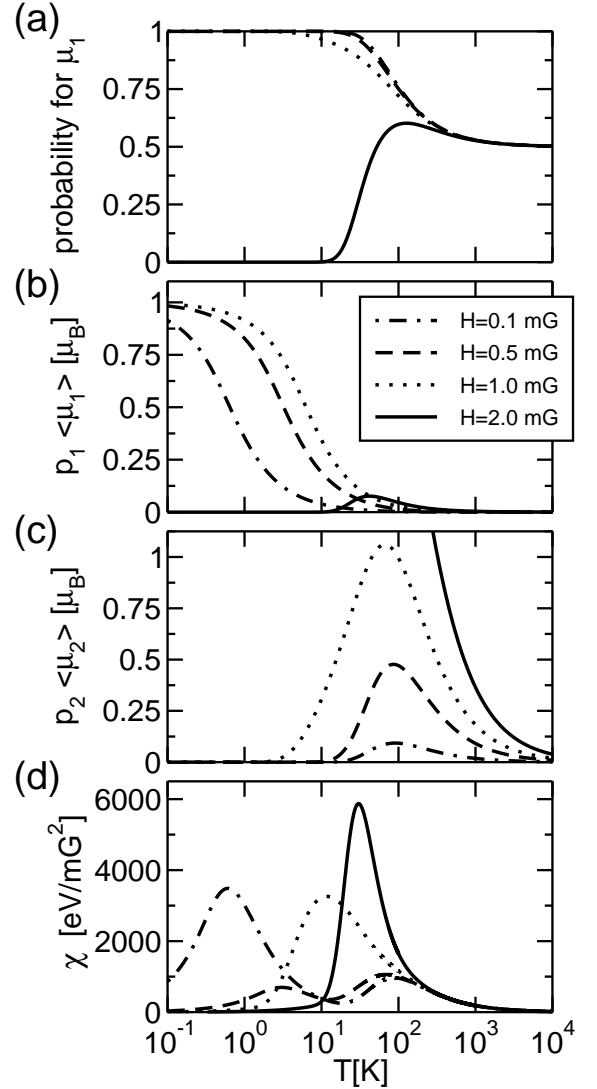}
\caption{Plots of physical properties versus temperature $T$ for $H=0.1$,
$0.5$, $1.0$, and $2.0~$mGauss. (a) Probability for isomer~1.  (b)
Contribution of isomer~1 and (c) of isomer~2 to the total average magnetic
moment $\muz$. (d) Magnetic susceptibility $\chi$.}
\label{fig:mu-prob-chi}
\end{figure}

Figure~\ref{fig:mu-prob-chi}(a) shows the occupation probability of isomer~1
($\mu_1=1\mu_B$) for different magnetic fields. At low magnetic fields, the
lower energetic isomer is predominantely occupied, while higher magnetic
fields lower the ground state energy of isomer~2 ($\mu_2=10\mu_B$) and
therefore the occupation is reversed. With increasing temperature
the probabilities become equal. The contributions of both isomers to the total
average magnetic moment $\muz$ are plotted in Fig.\ref{fig:mu-prob-chi}(b)
and (c). At low temperatures, small magnetic fields align the magnetic
moment of isomer~1 
$H$. With increasing temperature the mobility of the atoms is raised which
decreases the contribution of isomer~1 to $\muz$. 

From Fig.~\ref{fig:mu-cont} and Fig.~\ref{fig:mu-prob-chi} we can infer
that for temperatures $T\lesssim1$~K and an increasing magnetic field a
transition with a coexistence phase occurs (for coexistence in small
systems see~\cite{wales94,wales95,berry99,wales2000}). As the magnetic field is
raised the contributions of both isomers to $\muz$ are inverted which
causes a bimodal probability distribution $P(\mu_z)$ of the order
parameter $\mu_z$. The response of the system is observable as a hump in
the susceptibility $\chi=\partial_H \muz$ at about $30$~K for
$H=2.0$~mGauss, see Fig.~\ref{fig:mu-prob-chi}(d). 

\begin{figure*} 
\includegraphics[clip=,width=17.5cm]{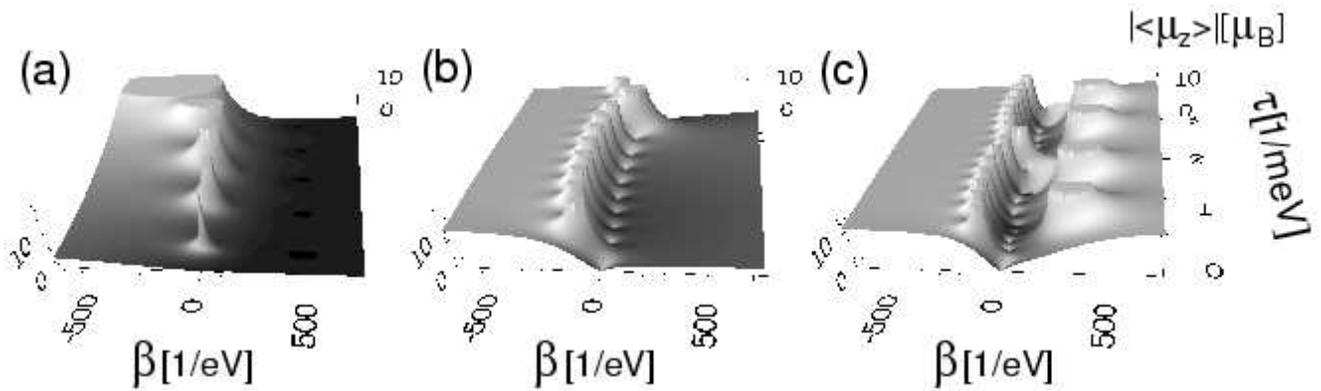}
\caption{The absolute value of the average magnetic moment
$\left|\muz\right|$ in the complex temperature plane for (a) 0.1mGauss,
(b) 1.2 mGauss, and (c) 2.0 mGauss.}
\label{fig:muc-zeros}
\end{figure*}

\begin{figure*}
\includegraphics[width=17cm]{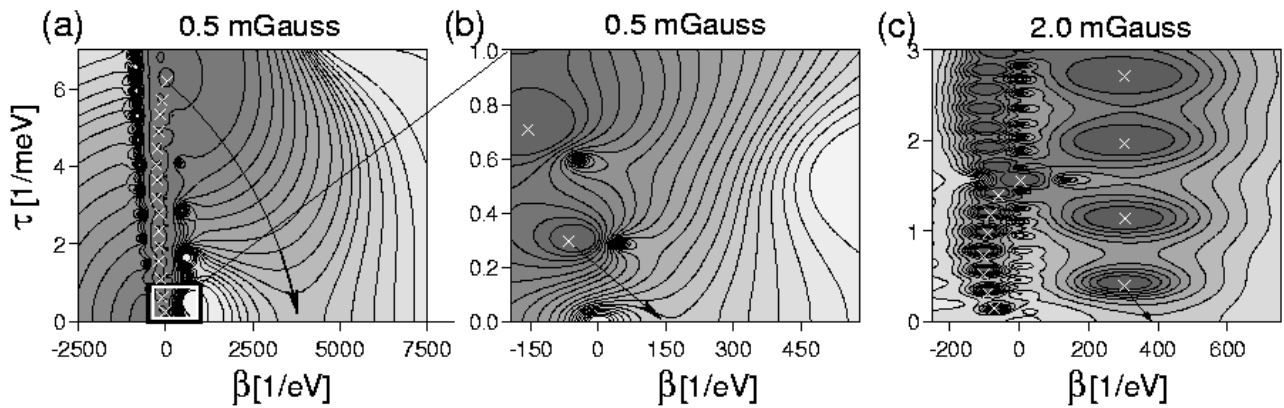}
\caption{Contour plot of the magnetic susceptibility $\chi$ over the
complex temperature plane, (a) for $H=0.5$~mGauss, (b) displays a
close-up of ca.~one tenth of this plane, and (c) for $H=2.0$~mGauss. The
white crosses indicate the location of the poles of $\chi$. The arrows are
a guide to the eye and visualize the ''radiation'' of the poles.}
\label{fig:chic-bw}
\end{figure*}

However, at temperatures about $10-1000$~K the situation is a bit more
complicated. With increasing temperature and at ``intermediate'' fields
($H\lesssim1.0~$mGauss) the contribution of isomer~1 decreases, while the
contribution of isomer~2 to $\muz$ for is raised up to
$1~\mu_B$. This also results in humps of the susceptibility $\chi$ but
$P(\mu_z)$ is unimodal because the magnetic moments of the $10\mu_B$
isomer are less aligned along the magnetic field. Since the first effect
can be regarded as pure magnetic field driven, it is not clear whether one
should attribute the humps for $H\lesssim1.0$~mGauss to a magnetic field
effect or to a temperature effect. The ``coexistence'' observable in the
contributions of both isomers to $\muz$ is fundamentally different, it
would be more correct to associate this with thermal excitation.  

Magnetic clusters are finite spin-systems. Such
systems, if their energy has an upper limit, can show an inverse change of
entropy with respect to energy corresponding to negative inverse
temperatures, $\beta=1/T=\partial_E S(E)=\partial_E \ln \Omega(E)$ with
$\Omega(E)$ being the density of states. Note, that negative temperatures
are attributed to the {\sl spin temperature} of the system.  Negative
temperatures have been measured, e.g.~in
LiF-crystals~\cite{Purcell:1951,Ramsey:1956,Landsberg:1959} by decoupling
the spin temperature from the kinetic energy contribution. The
spin-lattice relaxation time is large enough (up to hours) to assure that
the spin system is thermally stable and thus can come to
equilibrium~\cite{Abragam:1958}.

Obviously, the above presented indicators of phase transition thwart the
classification {\sl and} to some extent the distinction between different
phase transitions.  Within the microcanonical ensemble the occurence of
negative temperatures arises naturally because $T$ is an internal
parameter in contrast to the canonical ensemble. We consider the
positive {\sl and} negative inverse temperatures within the canonical
treatment to assure having sufficient information. 

Figure~\ref{fig:muc-zeros} shows
3-dimensional images of the distribution of zeros of $Z(\cbeta)$ in the
complex temperature plane. 
The poles of $|\muz|$ coincide with the zeros of the canonical
partition function $Z(\cbeta)$. 
For $H=0.1$~mGauss only one distribution of zeros lying at negative
temperatures is present (Fig.~\ref{fig:muc-zeros}(a)) corresponding to the
inverse occupation of the two isomers at negative temperatures. With
increasing magnetic field the shape of the distribution changes
(Fig.~\ref{fig:muc-zeros}(b)) and the influence of the poles of
$\left|\muz\right|$ on the real temperature axis becomes visible.
While this effect is hardly seen for
$H=0.1$~mGauss, one is not able to distinguish the isomers due to their
real-temperature values of $\muz$. For $H=1.2$~mGauss and $\beta>0$ the
average magnetic moment equals $1\mu_B$, whereas we have $\muz=10\mu_B$
for negative temperatures. At higher magnetic fields a second distribution
of zeros corresponding to the structural transition between both isomers
is seen.  

An inspection of both distributions of zeros reveals that the structural
transition is of first order, i.e., for the classification parameters we
have $\alpha=\gamma=0$. Whereas for the transition located at negative
temperatures we find $\alpha>0$ and $\gamma\neq0$, i.e., a second-order
transition, unless the magnetic field completely
disappears.

Another interesting example for first-order transitions, which become
second order with growing size have been found by Proykova et al.
\cite{proy1999}. There, two transitions have been found in TeF$_6$
clusters. They conclude that one of them is supposed to be of first-order,
which becomes continuous for larger clusters, because they find two minima
in the free energy as a function of an order parameter which merge to one
mimimum for larger cluster sizes.  The arbitrary choice of the order
parameter, however, might lead to different results
\cite{wales2000}.

Figure~\ref{fig:chic-bw}
displays the the magnetic susceptibility $\chi$ within the
complex temperature plane. The susceptibility is plotted versus
the complex inverse temperature. 
The blurred peaks of $\chi$ in Fig.~\ref{fig:mu-prob-chi}(d) can be clearly related to ``radiations'' of the zeros onto the
real axis. 

An inspection of the whole complex temperature plane reveals the
origin of the two maxima of the susceptibility at $H = 0.5$~mGauss on the
real axis (cmp. Fig.~\ref{fig:mu-prob-chi}(d)).  For $H=0.5$~mGauss the
hump in $\chi$ located at $T\approx3$~K ($\beta\approx3800$~1/eV) has its
origin in the distribution of
zeros lying at negative temperatures. Also the hump at $T=80$~K, where
earlier calculations have suggested that it is also related to a
structural transition~\cite{bdht96}, has its origin in this distribution
of zeros. 
For $2.0$~mGauss the
distribution of zeros lying at positive temperatures contributes most to
$\chi$ for real temperatures at $\beta\approx400$~1/eV~$\hat=\;
T\approx30$~K corresponding to the structural transition seen
in Fig.~\ref{fig:mu-prob-chi}(d). 

Since the distribution of poles of the considered thermodynamic function
(the distribution of zeros of $Z(\cbeta)$) is discrete the influences on
the real axis might be shielded by zeros of these function. For example, the
distributions of poles of $\muz$ and $\chi$ are surrounded by distributions
of zeros which are different for $\muz$ and $\chi$. These thermodynamic
functions are analytic everywhere except at the poles and zeros. Thus, if
there is a zero in the vicinity of a pole near the real axis the decrease
of $\muz$ or $\chi$ cannot be compensated.

In conclusion we have shown that the use of the complex inverse
temperature plane has advantages to common investigations of thermodynamic
functions. Within our classification scheme the order of a transition can
be clearly identified. By means of a simple two-isomer model for small
magnetic clusters we are able to identify two different types of phase
transitions. Furthermore, we found that signals in the magnetic
susceptibility at positive temperatures might have their origin at
negative temperatures. This also indicates that the inverse temperature is
analytic at $\beta=1/T=0$ and therefore should be used in calculations of
thermodynamic properties.

We thank E.~R.~Hilf for fruitful discussions and valuable comments.

\bibliographystyle{apsrev}

\end{document}